\documentclass[11pt,eqs]{article}
\usepackage{latexsym}

\def\cleq{\setcounter{equation}{0}}

\title{
T-dualization in a curved background in absence of a global symmetry
\thanks{Work supported in part by
the Serbian Ministry of Education, Science and Technological Development, under contract No. 171031.}}
\author{Lj. Davidovi\'c \thanks{e-mail: ljubica@ipb.ac.rs}
 and B. Sazdovi\'c \thanks{e-mail: sazdovic@ipb.ac.rs}\\
{\it Institute of Physics,}\\
{\it University of Belgrade,}\\
{\it 11001 Belgrade, P.O.Box 57, Serbia}}

\begin{document}
\maketitle
\begin{abstract}
We investigate T-duality of a closed string moving in a
weakly curved background of the second order.
A previously discussed weakly curved background
consisted of a flat metric and a linearly coordinate dependent
Kalb-Ramond field with an infinitesimal strength.
The background here considered differs from the above in
a coordinate dependent metric of the second order.
Consequently,
the corresponding Ricci tensor is nonzero.
As this background does not posses the
global shift symmetry the generalized  Buscher T-dualization procedure
is not applicable to it.
We redefine it and make it applicable to backgrounds without the global symmetry.
\end{abstract}

\section{Introduction}

T-duality \cite{PR,AAL,MAH,AS,sedam} was first introduced to represent the fact that
the toroidal compactifications \cite{BKS,KY,SS,LL} of a closed string
to a radius $R$ and a radius $1/R$ are equivalent.
Although the compactified theories are defined for different target spaces, their
spectrum is the same.
The observed symmetry lead to investigation for discovering  the connections
between the theories having the
same spectrum,
what resulted in T-dualization procedures.
These prescriptions
consist of rules which transform the given theory
to its T-dual theory.
The T-dual theories describing strings moving in a geometrically
different backgrounds,
having the same predictions, were found
for some particular backgrounds.
The general procedure, applicable to an arbitrary background is
still to be determined.

T-duality
was found to be connected with the isometries of a sigma model.
This discovery was included as an inevitable condition for
T-dualization, as the isometry was built in the T-dualization procedure.
The first procedure determining the T-duals of the constant background fields
was the Buscher procedure \cite{B,RV,AABL}.
This procedure, called the standard T-dualization procedure,
is founded in localizing the global isometry by introducing the
gauge fields, whose field strength is set to zero by a Lagrange multiplier term.
The gauged theory reduces to the initial theory for the equations of motion for the
Lagrange multiplier and to the T-dual theory for the equations of motion for the
gauge fields.
This procedure enabled investigation of the coordinate dependent backgrounds as well,
when the T-dualizations are performed along the directions
which do not appear as the background fields arguments.

The generalized T-dualization proposed in
paper \cite{OQ},
addressing the non-abelian isometries lead to observation
that
the application of T-dualization procedures  can lead to theories without an isometry.
These theories are obviously T-dual to the initial theories, but, the initial theories can not be obtained
as a result of the same T-dualization procedure starting with the theory without isometry.
This observation implied that T-duality in general must be understood from some other perspective.

The investigation of
the relations between non geometric backgrounds
lead to a new generalized T-dualization
in string field theory \cite{DH}.
It was proposed that particular non geometric backgrounds
should be understood as string backgrounds  which are the result of generalized T-du\-a\-li\-za\-tion
applied along nonisometry directions.
In paper \cite{Hull}, the conditions for a background to have a geometric or non geometric T-dual were sought for. It was concluded that the large class of sigma models that cannot be gauged can be T-dualized.

In paper \cite{EG},
one considers similarity transformations of
the stress energy tensor of a conformal field theory
which do not change the Virasoro alebra.
There exists the transformations of the background fields
which produce the same change of the stress tensor
as the change generated by some similarity transformations.
A particular generator of a similarity transformation
produces symmetry transformation, such as
general coordinate transformations and gauge transformations of a Kalb-Ramond field.
However some particular forms of these generators
produce T-duality transformations at critical radius.
Investigation leads to
T-dualization techniques directly applicable to an arbitrary string backgrounds.

In paper \cite{DS},
a generalization of a Buscher T-dualization procedure was given.
The generalized procedure is
applicable to backgrounds depending on all the space-time
coordinates,
along arbitrary background fields argument.
The procedure was realized for
a weakly curved background which consists of a constant metric and a coordinate dependent Kalb-Ramond field with an infinitesimal strength.
The difference between the generalized and standard Buscher procedure is in an
invariant coordinate.
Standardly, one
substitutes the derivatives with the covariant derivatives to obtain the gauge invariant action,
but in the generalized procedure
one additionally substitutes the argument of the background field with the
invariant argument.
It was confirmed in \cite{DS} that the generalized T-dualization procedure does not harm
the interchange of equations of motion and Bianchi identities \cite{BI}.
However, it strongly changes the geometry of a target space. The geometric space is transformed to a
double non-geometric space.
The commutative space is transformed to a non-commutative space, as shown in \cite{DNS}.
The closed string non-commutativity was previously investigated in \cite{NCM}.
The application of a procedure to an arbitrary set of coordinates was considered in \cite{DS1}.
It was concluded that the geometric background again transforms to a double space, with double coordinates present for both T-dualized and undualized directions.

In the present paper,
we consider the weakly curved background of the second order.
We take a metric which consists of a constant and quadratic in coordinate term
and linearly coordinate dependent Kalb-Ramond field.
This background does not posses the global shift symmetry.
In comparison to
the previously considered backgrounds there is an additional difference.
The Ricci tensor of the metric here considered is nonzero.
The background has to be the solution of the space-time equations of motion,
obtained from the demand of the conformal invariance of the
quantum theory.
To satisfy these equations one takes
the coordinate dependent parts to be infinitesimal.

The original form of the generalized Buscher T-dualization procedure \cite{DS} is not applicable to
 a weakly curved background of the second order.
Here, we search for the procedure which will be applicable and will preserve the
general features of the previous procedure.
We find the appropriate formulation and investigate the properties and the consequences of the new
generalization.
We apply the procedure along all space-time coordinates and obtain the T-dual theory.
We obtain a geometrical structure that differs  from the double non geometrical space.
The dual background field arguments do not depend only on the dual coordinate and its double.
However, the application of the procedure to all dual coordinates
leads again to the initial theory.
We obtain T-dual coordinate transformation laws and confirm that T-duality interchanges equations of motion and Bianchi identities.

\section{Bosonic string action and choice of background}
\cleq

Let us
consider the closed bosonic string propagating in the background fields:
a metric $G_{\mu\nu}$ and a Kalb-Ramond antisymmetric tensor field $B_{\mu\nu}$,
described by the action
\begin{equation}\label{eq:actiono}
S[x] = \kappa \int_{\Sigma} d^2\xi\,\partial_{\alpha}x^{\mu}
\Big[\frac{1}{2}\eta^{\alpha\beta}G_{\mu\nu}(x)
+{\epsilon^{\alpha\beta}}B_{\mu\nu}(x)\Big]
\partial_{\beta}x^{\nu},
\quad \varepsilon^{01}=-1.
\end{equation}
The integration goes over two-dimensional world-sheet $\Sigma$
parametrized by $\xi^{\alpha}\ ,\alpha=0,1$ ($\xi^{0}=\tau$, $\xi^{1}=\sigma$).
The coordinates of the D-dimensional space-time
are marked by $x^{\mu}(\xi),\
\mu=0,1,...,D-1$.
From the action principle one obtains the equations of motion
\begin{equation}\label{eq:motion}
{\ddot{x}}^\mu-x''^\mu+2B^\mu_{\ \nu\rho}{\dot{x}}^\nu x'^\rho
+\Gamma^\mu_{\nu\rho}({\dot{x}}^\nu{\dot{x}}^\rho-x'^\nu x'^\rho)=0,
\end{equation}
where
$B^\mu_{\ \nu\rho}=(G^{-1})^{\mu\sigma}B_{\sigma\nu\rho}$ and
$B_{\mu\nu\rho}=\partial_\mu B_{\nu\rho}+\partial_\nu B_{\rho\mu}+\partial_\rho B_{\mu\nu}$
is the field strength of the field $B_{\mu \nu}$ and
$\Gamma^\mu_{\nu\rho}
=\frac{1}{2}(G^{-1})^{\mu\sigma}(\partial_\nu G_{\rho\sigma}+\partial_\rho G_{\sigma\nu}
-\partial_\sigma G_{\nu\rho})$
is a Christoffel symbol.

Introducing the light-cone coordinates and their derivatives
\begin{equation}
\xi^{\pm}=\frac{1}{2}(\tau\pm\sigma),
\qquad
\partial_{\pm}=
\partial_{\tau}\pm\partial_{\sigma},
\end{equation}
the action (\ref{eq:actiono}) can be rewritten as
\begin{equation}\label{eq:action1}
S[x] = \kappa \int_{\Sigma} d^2\xi\
\partial_{+}x^{\mu}
\Pi_{+\mu\nu}(x)
\partial_{-}x^{\nu},
\end{equation}
where $\Pi_{\pm\mu\nu}$ is
the combination of background fields, defined by
\begin{eqnarray}\label{eq:pi}
\Pi_{\pm\mu\nu}(x)=
B_{\mu\nu}(x)
\pm\frac{1}{2}G_{\mu\nu}(x).
\end{eqnarray}
The equation of motion (\ref{eq:motion}) can be rewritten as
\begin{equation}\label{eq:motion1}
\partial_{+}\partial_{-}x^\mu
+\Big{(}\Gamma^{\mu}_{\nu\rho}-B^{\mu}_{\ \nu\rho}\Big{)}
\partial_{+}x^\nu\partial_{-}x^\rho=0.
\end{equation}

In order to obtain a conformally invariant quantum theory,
the background fields must obey the space-time equations of motion,
which for the constant dilaton field have the following form
\begin{equation}\label{eq:ste}
R_{\mu \nu} - B_{\mu \rho \sigma}
B_{\nu}^{\ \rho \sigma}=0\, ,
\end{equation}
\begin{equation}\label{eq:ste1}
D_\rho B^{\rho}_{\ \mu \nu} = 0,
\end{equation}
where $R_{\mu \nu}$ is Ricci tensor defined by
\begin{equation}\label{eq:riman}
R_{\mu \nu}=R^{\rho}_{\ \mu\rho\nu},\quad
R^{\rho}_{\ \mu\sigma\nu}=
\Gamma^{\rho}_{\mu\nu,\sigma}-\Gamma^{\rho}_{\mu\sigma,\nu}
+\Gamma^{\tau}_{\mu\nu}\Gamma^{\rho}_{\tau\sigma}
-\Gamma^{\tau}_{\mu\sigma}\Gamma^{\rho}_{\tau\nu},
\end{equation}
and $D_\mu$ is a covariant derivative
\begin{equation}\label{eq:beta1exp}
D_\rho B^{\sigma}_{\ \mu \nu}=
\partial_\rho B^{\sigma}_{\ \mu \nu}
+\Gamma^{\sigma}_{\varepsilon\rho} B^{\varepsilon}_{\ \mu \nu}
-\Gamma^{\varepsilon}_{\mu\rho}B^{\sigma}_{\ \varepsilon \nu}
-\Gamma^{\varepsilon}_{\nu\rho}B^{\sigma}_{\ \mu\varepsilon}.
\end{equation}
We will consider the following solution of the space-time equations
of motion (\ref{eq:ste}), (\ref{eq:ste1})
\begin{equation}\label{eq:wcb}
G_{\mu\nu}(x)=g_{\mu\nu}+3h^{2}_{\mu\nu}(x),
\quad B_{\mu\nu}(x)=b_{\mu\nu}+h_{\mu\nu}(x),
\end{equation}
with
$g_{\mu\nu},b_{\mu\nu}=const$ and
$h_{\mu\nu}\equiv\frac{1}{3}B_{\mu\nu\rho}x^\rho,\quad
h^{2}_{\mu\nu}\equiv(hg^{-1}h)_{\mu\nu},\,$
where $B_{\mu\nu\rho}$ is constant and infinitesimal.
Throughout the paper the calculation will be done
up to the
second order in $B_{\mu\nu\rho}$.
We will refer to this solution as
the weakly curved background of the second order.
More general discussion about the solutions of the space-time equations of motion,
up to the second order terms, can be found in Ref.
\cite{SC}.

Let us demonstrate that (\ref{eq:wcb}) satisfies (\ref{eq:ste})
and (\ref{eq:ste1}).
The inverse metric and the Cristoffel symbol are
\begin{equation}
(G^{-1})^{\mu\nu}=(g^{-1})^{\mu\nu}
-3(h^{2})^{\mu\nu},
\end{equation}
\begin{equation}\label{eq:chris}
\Gamma^\mu_{\nu\rho}=-
B^{\mu}_{\ \nu\sigma}h^{\sigma}_{\ \rho}
-B^{\mu}_{\ \rho\sigma}h^{\sigma}_{\ \nu},
\end{equation}
with $(h^{2})^{\mu\nu}=(g^{-1}hg^{-1}hg^{-1})^{\mu\nu}$ and $B^{\mu}_{\ \nu\sigma}h^{\sigma}_{\ \rho}=
(g^{-1})^{\mu\varepsilon}B_{\varepsilon\nu\sigma}(g^{-1})^{\sigma\tau}h_{\tau\rho}$.
Therefore, the Ricci tensor equals
\begin{equation}
R_{\mu\nu}=B_{\mu\rho\sigma}B^{\rho\sigma}_{\ \ \nu},
\end{equation}
which is just the eq. (\ref{eq:ste}).
The equation (\ref{eq:ste1}) is satisfied,
because the term corresponding to the first term in (\ref{eq:beta1exp})
is zero and the others can be neglected as the
third order terms in $B_{\mu\nu\rho}$.

Let us notice, using (\ref{eq:riman}) and (\ref{eq:chris}), that the coefficient in the second order term of the metric (\ref{eq:wcb})
is in fact the Riemann curvature tensor
\begin{equation}
G_{\mu\nu}=g_{\mu\nu}-\frac{1}{3}R_{\mu\rho\nu\sigma}x^\rho x^\sigma.
\end{equation}

The solutions which  where previously investigated in this context where the constant background
and the weakly curved background of the first
order.
In both cases
the  Ricci tensor $R_{\mu \nu}$ is absent,
in the first case because it equals zero and in the second
because it is neglected as the second order term. Here, the Ricci tensor  is
of the second order and its contribution
becomes nontrivial
because we work up to the second order in $B_{\mu\nu\rho}$.

\section{T-dualization procedure}\label{sec:tproc}
\cleq

In the majority of papers addressing T-dualization of a bosonic string theory,
one performs T-dualizations
along directions on which the background does not depend.
The first procedure, applicable to
coordinates on which the background fields depend,
the generalization of the Buscher T-dualization procedure, was presented in \cite{DS}.
It was applied to a bosonic string moving in the weakly curved background, composed of a constant metric and
a linearly coordinate dependent Kalb-Ramond field with an infinitesimal strength. This theory has a global
shift symmetry. This fact is used in the T-dualization prescription,
which relies on gauging the global symmetry.
The locally invariant action was built substituting the
ordinary derivatives with the covariant ones
and substituting
 the coordinate in the argument of the background fields with the
invariant coordinate (a line integral of a covariant derivatives of the original coordinates).
The physical equivalence was achieved by introduction of the Lagrange multiplier term,
which makes the gauge fields nonphysical.

\subsection{Auxiliary action}

Here, we will consider the weakly curved background of the second order.
As in the first order weakly curved background,
the Kalb-Ramond field is linear in coordinate and has an infinitesimal field strength.
The metric however, beside of a constant term has a quadratic in coordinate part which is an infinitesimal of the second order.
Such a metric has an infinitesimal but nonzero Ricci tensor $R_{\mu\nu}\neq 0$.
The bosonic string theory in this background does not possess the shift symmetry.
However, defining of the new T-dualization rules on the grounds of the existing procedure is still possible.
The main object in the conventional procedure, is the gauge fixed action which reduces
to the initial action for the equations of motion for the Lagrange multipliers
and becomes T-dual action for the equations of motion for the gauge fields.
Here we will define its substitution, which inherits these two features.
We postulate the auxiliary action by
\begin{equation}\label{eq:spom}
S_{aux}[y,v_{\pm}]=\kappa\int d^{2}\xi\Big{[}
v_{+}^\mu \Pi_{+\mu\nu}\big(\Delta V\big) v_{-}^\nu
+\frac{1}{2}(v^{\mu}_{+}\partial_{-}y_\mu
-v^{\mu}_{-}\partial_{+}y_\mu)
\Big{]}.
\end{equation}
It can be obtained from the initial action (\ref{eq:action1}),
by making the following substitutions
\begin{eqnarray}
\partial_{\pm}x^\mu\rightarrow v^{\mu}_{\pm},
&&x^\mu\rightarrow\Delta V^\mu
\end{eqnarray}
and adding the Lagrange multiplier  $y_\mu$ term.
This action is of the same form as the gauge fixed action,
however, $v^{\mu}_{\pm}$ are here some auxiliary fields,
which take over the role of the gauge fields.
Similarly as in \cite{DS}, the argument of the background fields
is the line integral of the auxiliary fields
taken along a path $P$ (from $\xi_{0}$ to $\xi$)
\begin{equation}\label{eq:vdef}
\Delta V^\mu[v_{+},v_{-}]\equiv
\int_{P}d\xi^\alpha v^{\mu}_{\alpha}
=\int_{P}(d\xi^{+} v^{\mu}_{+}
+d\xi^{-} v^{\mu}_{-}).
\end{equation}
Note that as well as in Ref. \cite{DS},
the equation of motion with respect to $y_\mu$
forces the "field strength" to vanish $\partial_{+}v_{-}^\mu-\partial_{-}v_{+}^\mu=0$,
which is just the condition for the path independence of $\Delta V^\mu$.
In the resulting theories, the argument reduces to $\Delta V^\mu=V^\mu(\xi)-V^\mu(\xi_{0})$ and we will chose the value of $V^\mu(\xi_{0})$ to be zero.

\subsection{
From the auxiliary to the initial and T-dual theory
}

Let us confirm that the auxiliary action (\ref{eq:spom})
becomes the initial action (\ref{eq:action1}) for
the equations of motion obtained varying over
the Lagrange multiplier $y_\mu$
\begin{equation}\label{eq:emy}
\partial_{+}v^{\mu}_{-}
-\partial_{-}v^{\mu}_{+}=0.
\end{equation}
Using their solution
\begin{equation}\label{eq:vsol}
v^{\mu}_{\pm}=\partial_{\pm}x^\mu,
\end{equation}
one obtains
$V^\mu(\xi)=x^\mu(\xi)$,
and therefore taking $x^\mu(\xi_{0})=0$
the auxiliary action reduces to the initial action (\ref{eq:action1}).

The equations of motion for the auxiliary fields $v^\mu_{\pm}$ are
\begin{equation}\label{eq:emv}
\Pi_{\mp\mu\nu}(V)v_{\pm}^{\nu}
+\frac{1}{2}\partial_{\pm}y_\mu=
\mp\beta^{\mp}_\mu(V).
\end{equation}
Here the functions $\beta^\pm_\mu$ are defined by
\begin{equation}\label{eq:betadef}
\delta_{V}S_{aux}=-\kappa\int
d\xi^{2}
\beta^\alpha_\mu\delta v^\mu_\alpha,
\end{equation}
where $\delta_{V}S_{aux}$ stands for the variation of
the action (\ref{eq:spom}) over the background field argument $V^\mu$.

Let us introduce
the following background fields:
an effective metric and a non-commu\-ta\-ti\-vi\-ty parameter,
defined by
\begin{equation}
G^{E}_{\mu\nu}\equiv\big(G-4BG^{-1}B\big)_{\mu\nu},
\quad
\theta^{\mu\nu}\equiv
-\frac{2}{\kappa}
(G^{-1}_{E}BG^{-1})^{\mu\nu},
\end{equation}
and their combinations
\begin{eqnarray}
{\Theta}^{\mu\nu}_{\pm}
=-\frac{2}{\kappa}
(G^{-1}_{E}
\Pi_{\pm}
G^{-1})^{\mu\nu}
=\theta^{\mu\nu}
\mp\frac{1}{\kappa}(G_{E}^{-1})^{\mu\nu},
\end{eqnarray}
which are the inverses of the background field compositions $2\kappa\Pi_{\mp\mu\nu}$.
Now,
one can rewrite the equations of motion (\ref{eq:emv}) as
\begin{eqnarray}\label{eq:vsol1}
v_{\pm}^{\mu}(y)&=&
-\kappa\,{\Theta}^{\mu\nu}_{\pm}\big(V(y)\big)
\Big{[}
\partial_{\pm}y_\nu
\pm2\beta^{\mp}_\nu\big(V(y)\big)
\Big{]}.
\end{eqnarray}
The equation (\ref{eq:vsol1}) is not the solution
of (\ref{eq:emv}), because $v^\mu_{\pm}$ appears within the argument
$ V^\mu$  of both $\Theta_\pm^{\mu\nu}$ and $\beta^{\mp}_{\mu}$.
We will solve this equation iteratively.

The T-dual theory is obtained, by inserting the equation of motion (\ref{eq:vsol1}) into the action (\ref{eq:spom})
\begin{equation}\label{eq:tdualaction}
^\star S[y,v_{\pm}]=
\frac{\kappa^{2}}{2}
\int d^{2}\xi\
\Big{[}
\partial_{+}y_\mu\Theta_{-}^{\mu\nu}\big(V(y)\big)
\partial_{-}y_\nu
+4\beta^{-}_{\mu}\big(V(y)\big)
\Theta_{-}^{\mu\nu}\big(V(y)\big)
\beta^{+}_{\nu}\big( V(y)\big)
\Big{]}.
\end{equation}
In order to obtain the explicit form of the T-dual action one has to calculate
the beta functions $\beta^{\pm}_{\mu}$ for a concrete background,
solve (\ref{eq:vsol1}) to find the explicit $y$-dependence of the auxiliary fields
$v^\mu_{\pm}=v^\mu_{\pm}(y)$,
and therefrom determine the argument of the dual background fields $V^\mu(y)$.

\section{T-dual action in a weakly curved background}
\cleq

Let us find the explicit expression for the T-dual action (\ref{eq:tdualaction}),
in the weakly curved background of the second order.
The main task is to obtain the $\beta^{\pm}_{\mu}$ functions (\ref{eq:betaexp}), which are calculated in appendix \ref{sec:betaj}.
Because they are infinitesimal,
it is enough to consider their first order value and to determine the zeroth order value of $V^\mu$,
in order to calculate the last term in the action
\begin{equation}\label{eq:actionb}
4\beta^{-}_{1\mu}\big( V_{0}(y)\big)
\Theta_{0-}^{\mu\nu}\beta^{+}_{1\nu}\big( V_{0}(y)\big)=
\partial_{+}V_{0}^\mu
h_{\mu\nu}\big( V_{0}(y)\big)\Theta^{\nu\rho}_{0-}h_{\rho\sigma}\big( V_{0}(y)\big)
\partial_{-}V_{0}^\sigma.
\end{equation}

Let us  find the explicit form of the dual background fields argument.
We will
solve the equations (\ref{eq:vsol1}) iteratively and find the
zeroth and the first order in $B_{\mu\nu\rho}$ values of
the auxiliary fields $v_{\pm}^{\mu}(y)$.
In the zeroth order
one has
\begin{equation}\label{eq:vo}
v_{0\pm}^{\mu}(y)=
-\kappa\,{\Theta}^{\mu\nu}_{0\pm}\,
\partial_{\pm}y_\nu,
\end{equation}
consequently the zeroth order value of $ V^\mu$ defined in (\ref{eq:vdef}) is
\begin{eqnarray}\label{eq:vnula}
 V_{0}^{\mu}
&=&
-\frac{\kappa}{2}\big(
\Theta^{\mu\nu}_{0+}+\Theta^{\mu\nu}_{0-}
\big)y^{(0)}_\nu
-\frac{\kappa}{2}\big(
\Theta^{\mu\nu}_{0+}-\Theta^{\mu\nu}_{0-}
\big){\tilde{y}}^{(0)}_\nu
\nonumber\\
&=&
-\kappa\theta_{0}^{\mu\nu}
y_\nu^{(0)}
+
(g^{-1}_{E})^{\mu\nu}
{\tilde{y}}_\nu^{(0)},
\end{eqnarray}
where ${\tilde{y}}_\mu$ is
a double coordinate defined by
\begin{equation}\label{eq:y}
{\tilde{y}}_\mu\equiv
\int_{P} d\xi^\alpha \varepsilon^\beta_{\ \alpha}\partial_{\beta}y_\mu.
\end{equation}

Now, using (\ref{eq:vnula}) and (\ref{eq:teta}), the last term (\ref{eq:actionb})
 in the T-dual action (\ref{eq:tdualaction}),  becomes
\begin{eqnarray}\label{eq:actionb1}
4\beta^{-}_{1\mu}\big(V_{0}(y)\big)
\Theta_{0-}^{\mu\nu}\beta^{+}_{1\nu}\big(V_{0}(y)\big)&=&
-\,\frac{\kappa}{2}
\partial_{+}y_\mu
\Big{[}\Theta_{1-}\big(V_{0}(y)\big)\Pi_{0+}\Theta_{1-}\big(V_{0}(y)\big)\Big{]}^{\mu\nu}
\partial_{-}y_\nu
\nonumber\\
&\equiv&
-\,\partial_{+}y_\mu\,\Delta^{\mu\nu}_{-}\big(V_{0}(y)\big)
\partial_{-}y_\nu,
\end{eqnarray}
with $\Delta^{\mu\nu}_\pm$ explicitly given by (\ref{eq:delta}).

The first order value of the auxiliary field $v_\pm^\mu$, defined by (\ref{eq:vsol1}),
is obtained using
the first order value of $\beta^{\pm}_{\mu}$, (\ref{eq:betaexp}),
and
the expressions for $\Theta^{\mu\nu}_{1\pm}$
given by (\ref{eq:teta})
\begin{equation}\label{eq:v1}
v_{1\pm}^{\mu}(y)=
-\kappa\,{\Theta}^{\mu\nu}_{0\pm}
\partial_{\pm}y^{(1)}_\nu
-\frac{3\kappa}{2}
\Theta^{\mu\nu}_{1\pm}(V_{0}(y))
\partial_{\pm}y_\nu^{(0)}.
\end{equation}
Let us note that, because of (\ref{eq:vo}) and (\ref{eq:v1}), the complete first order value of the auxiliary field can be written as
\begin{equation}\label{eq:vdt}
v_{\pm}^{(1)\mu}(y)=-\kappa\,^\diamond\Theta^{(1)\mu\nu}_\pm(V(y))\partial_\pm y_\nu,
\end{equation}
where $^\diamond\Theta^{\mu\nu}_\pm$ is defined in (\ref{eq:dteta}).

Substituting (\ref{eq:v1}) into (\ref{eq:vdef}), we obtain
the first order value of $V^\mu$
\begin{eqnarray}\label{eq:vjedan}
&&V_{1}^\mu(y)=
-\frac{\kappa}{2}\big(
\Theta^{\mu\nu}_{0+}+\Theta^{\mu\nu}_{0-}
\big)y^{(1)}_\nu
-\frac{\kappa}{2}\big(
\Theta^{\mu\nu}_{0+}-\Theta^{\mu\nu}_{0-}
\big){\tilde{y}}^{(1)}_\nu
\nonumber\\
&&\quad-\frac{\kappa^{3}}{2}\big(
\Theta^{\mu\nu}_{0+}+\Theta^{\mu\nu}_{0-}
\big)B_{\nu\rho\sigma}
\Big(
\Theta^{\sigma\tau}_{0+}\Theta^{\rho\varepsilon}_{0+}M_{+\tau\varepsilon}(y)
+\Theta^{\sigma\tau}_{0-}\Theta^{\rho\varepsilon}_{0-}M_{-\tau\varepsilon}(y)
+\Theta^{\sigma\tau}_{0-}\Theta^{\rho\varepsilon}_{0+}
\widetilde{(y^{(0)}_{-\tau}y^{(0)}_{+\varepsilon})}
\Big)
\nonumber\\
&&\quad-\frac{\kappa^{3}}{2}\big(
\Theta^{\mu\nu}_{0+}-\Theta^{\mu\nu}_{0-}
\big)B_{\nu\rho\sigma}
\Big(
\Theta^{\sigma\tau}_{0+}\Theta^{\rho\varepsilon}_{0+}M_{+\tau\varepsilon}(y)
-\Theta^{\sigma\tau}_{0-}\Theta^{\rho\varepsilon}_{0-}M_{-\tau\varepsilon}(y)
+\Theta^{\sigma\tau}_{0-}\Theta^{\rho\varepsilon}_{0+}\,y^{(0)}_{-\tau}y^{(0)}_{+\varepsilon}
\Big),
\nonumber\\
\end{eqnarray}
where
\begin{equation}\label{eq:moment}
M_{\pm\mu\nu}(y)\equiv
\frac{1}{2}
\int d\xi^\pm\,\Big( y^{(0)}_{\pm\mu}\partial_\pm y^{(0)}_{\pm\nu}
-y^{(0)}_{\pm\nu}\partial_\pm y^{(0)}_{\pm\mu}
\Big),
\end{equation}
and $\widetilde{(y^{(0)}_{-\tau}y^{(0)}_{+\varepsilon})}$ is a double (defined by (\ref{eq:y})) of the quantity
$y^{(0)}_{-\tau}y^{(0)}_{+\varepsilon}$.

Once the
argument of the background fields is calculated,
we can write the explicit form of
the T-dual action
\begin{eqnarray}\label{eq:tdualaction1}
^\star S[y]&=&
\frac{\kappa^{2}}{2}
\int d^{2}\xi\
\partial_{+}y_\mu
\,^\dag
\Theta_{-}^{\mu\nu}\big(V(y)\big)
\partial_{-}y_\nu
\nonumber\\
&\equiv&
\frac{\kappa^{2}}{2}
\int d^{2}\xi\
\partial_{+}y_\mu
\Big{[}
\Theta_{-}^{\mu\nu}\big(V(y)\big)
-\Delta^{\mu\nu}_{-}\big(V_{0}(y)\big)
\Big{]}
\partial_{-}y_\nu,
\end{eqnarray}
with $V^\mu=V^{(1)\mu}=V^\mu_{0}+V^\mu_{1}$ given by (\ref{eq:vnula})
and (\ref{eq:vjedan}).
The second term in the dual background fields composition $\Delta^{\mu\nu}_{-}$ is
the contribution from the term quadratic in $\beta$,
and has a form (\ref{eq:delta}).

Comparing the initial action (\ref{eq:action1}) with the T-dual action (\ref{eq:tdualaction1}),
one can conclude that they are equal under the following transformations
$$
\partial_{\pm} x^\mu\rightarrow\partial_{\pm}y_\mu,
$$
\begin{eqnarray}\label{eq:tetadualna}
\Pi_{+\mu\nu}(x)\rightarrow
\,^\star\Pi_{+}^{\mu\nu}(y)&=&
\frac{\kappa}{2}\Big(
\Theta_{-}^{\mu\nu}\big(V(y)\big)
-\Delta^{\mu\nu}_{-}\big(V_{0}(y)\big)
\Big)
\nonumber\\
&\equiv&\frac{\kappa}{2}\,^\dag\Theta_{-}^{\mu\nu}\big(V(y)\big).
\end{eqnarray}
The T-dual metric (the symmetric part of the T-dual background fields composition) and the T-dual Kalb-Ramond field
(the antisymmetric part)
 are
\begin{eqnarray}\label{eq:dm}
^\star G^{\mu\nu}=
(G^{-1}_{E})^{\mu\nu}
-\kappa^{2}(
\theta_{0}h\theta_{0}hg^{-1}_{E}
+\theta_{0}hg^{-1}_{E}h\theta_{0}
+g^{-1}_{E}h\theta_{0}h\theta_{0}
)^{\mu\nu}
-(g^{-1}_{E}hg^{-1}_{E}hg^{-1}_{E})^{\mu\nu}\!,
\end{eqnarray}
and
\begin{eqnarray}
^\star B^{\mu\nu}=
\frac{\kappa}{2}\theta^{\mu\nu}
-\frac{\kappa^{3}}{2}(\theta_{0}h\theta_{0}h\theta_{0})^{\mu\nu}
-\frac{\kappa}{2}(
\theta_{0}hg^{-1}_{E}hg^{-1}_{E}
+g^{-1}_{E}h\theta_{0}hg^{-1}_{E}
+g^{-1}_{E}hg^{-1}_{E}h\theta_{0}
)^{\mu\nu}\!.
\end{eqnarray}

The characteristics of the dual geometry, are considered in section \ref{sec:uporedjenje}.

\section{T-dual of T-dual}\label{sec:dualdual}
\cleq

Let us now follow the prescription of Sec.\ref{sec:tproc},
and show that the T-dual of a T-dual theory is the original theory.
To obtain the auxiliary action of the T-dual action (\ref{eq:tdualaction1}),
let us substitute the dual coordinate derivatives $\partial_{\pm}y_\mu$
with some auxiliary fields $u_{\pm\mu}$,
substitute the coordinate in the argument of background fields with
$\Delta U_\mu=\int (d\xi^{+}u_{+\mu}+d\xi^{-}u_{-\mu})$ and
require the ''flatness`` of $u_{\pm\mu}$
by introducing the Lagrange multiplier terms
\begin{equation}\label{eq:sdualaux}
^\star S_{aux}[z,u_\pm]=
\frac{\kappa}{2}
\int d^{2}\xi\Big{[}
\kappa u_{+\mu}\,^\dag\Theta_{-}^{\mu\nu}\big(V(\Delta U)\big)\,u_{-\nu}
+u_{+\mu}\partial_{-}z^\mu
-u_{-\mu}\partial_{+}z^\mu
\Big{]}.
\end{equation}
For the solution
$u_{\pm\mu}=\partial_{\pm}y_\mu$
of the equations of motion
$\partial_{-}u_{+\mu}+\partial_{+}u_{-\mu}=0$,
which are
 obtained varying the auxiliary action over the Lagrange multiplier $z^\mu$,
the variable $U_\mu$ reduces to $y_\mu$ ($y_\mu(\xi_{0})=0$).
So, the auxiliary action reduces to the original one.

The original theory should be obtained for
the equations of motion for the auxiliary fields $u_{\pm\mu}$
\begin{equation}
\kappa\,{^\dag\Theta}^{\mu\nu}_{\pm}\big(V(U)\big)u_{\pm\nu}
+\partial_{\pm}z^\mu=
\mp\kappa\,{^\star\beta}^{\mp\mu}\big(V(U)\big),
\end{equation}
with $^\star\beta^{\mp\mu}$ defined  by (\ref{eq:betadualdef}),
being the contribution from the variation over the background fields argument.
Multiplying the equations by $^\dag\Pi_{\mp\mu\nu}$
(the inverse of the background field composition
$^\dag\Theta^{\mu\nu}_{\pm}$,
defined in (\ref{eq:pik})) one obtains
\begin{equation}\label{eq:ue}
u_{\pm\mu}
=-2\kappa\,^\dag\Pi_{\mp\mu\nu}\big(V(U)\big)\Big[
\frac{1}{\kappa}\partial_{\pm}z^\nu
\pm{^\star \beta}^{\mp\nu}\big(V(U)\big)\Big].
\end{equation}
Using the last equation and the first order value of the dual beta function $^\star\beta^{\mp\nu}$ given by (\ref{eq:veza}), we can determine the value of the variable $U_\mu$,
up to the first order
\begin{eqnarray}\label{eq:uexp}
U^{(1)}_\mu(z)=
-2b_{\mu\nu}z^\nu+g_{\mu\nu}\tilde{z}^\nu
-B_{\mu\nu\rho}
\Big[
M^{\rho\nu}_{+}(z^{(0)})+M^{\rho\nu}_{-}(z^{(0)})-\widetilde{(z^{(0)\rho}_{+} z^{(0)\nu}_{-})}
\Big],
\end{eqnarray}
with $M_\pm^{\mu\nu}(z)$ defined in (\ref{eq:moment}).
Substituting (\ref{eq:uexp}) to (\ref{eq:vnula}) and (\ref{eq:vjedan}) we confirm that $V^\mu(U)=z^\mu$.

So, substituting (\ref{eq:ue}) to the action (\ref{eq:sdualaux}), we obtain
\begin{equation}\label{eq:dejrez}
{^\star}{^\star}S_{aux}[z]=
\kappa
\int d^{2}\xi\Big{[}
\partial_{+}z^\mu\,^\dag\Pi_{+\mu\nu}(z)
\partial_{-}z^\nu
+\kappa^{2}\,
{^\star\beta}^{-\mu}(z)\,^\dag\Pi_{+\mu\nu}(z)
{^\star\beta}^{+\nu}(z)
\Big{]}.
\end{equation}
Using the first order value of $^\star\beta^{\pm\mu}$, given by (\ref{eq:veza}),
the second term of the action becomes
$-2\kappa
\partial_{+}z^\mu
(\Pi_{0+}\Delta_{-}(z)\Pi_{0+})_{\mu\nu}
\partial_{-}z^\nu
$
and therefore the action (\ref{eq:dejrez}) is just the initial action (\ref{eq:action1})
\begin{equation}
{^\star}{^\star}S_{aux}[z]=
\kappa
\int d^{2}\xi\,
\partial_{+}z^\mu\,\Pi_{+\mu\nu}(z)
\partial_{-}z^\nu.
\end{equation}

\section{Features of T-duality}\label{sec:bjanki}
\cleq

In the previous sections, we showed how the original and its T-dual theory can be transformed one to
the other. In both directions, both theories follow from the auxiliary action and are obtained for a concrete form of the auxiliary fields. Comparing these auxiliary fields one obtains the T-dual coordinate transformation laws.

In section \ref{sec:tproc}, we showed how the original theory can be transformed into its T-dual theory.
So, comparing
the expressions for the auxiliary fields (\ref{eq:vsol}) and (\ref{eq:vsol1}),
one obtains the T-dual coordinate transformation law
\begin{equation}\label{eq:zapr}
\partial_{\pm}x^\mu
\cong -\kappa\,{\Theta}^{\mu\nu}_{\pm}\big( V(y)\big)
\Big{[}
\partial_{\pm}y_\nu
\pm2\beta^{\mp}_\nu\big( V(y)\big)
\Big{]}.
\end{equation}
In the first order this law implies
\begin{equation}\label{eq:xV}
x^{(1)\mu}\cong V^{(1)\mu}(y).
\end{equation}
Substituting the beta functions (\ref{eq:betaexp}) into the transformation law (\ref{eq:zapr}),
we obtain
\begin{equation}\label{eq:zakon}
\partial_{\pm} x^\mu\cong -\kappa\,
\Big[\,
{^\diamond\Theta}^{\mu\nu}_\pm\big(V(y)\big)+\Delta^{\mu\nu}_\pm\big(V(y)\big)
\Big]
\partial_\pm y_\nu,
\end{equation}
with  $^\diamond\Theta^{\mu\nu}_\pm$ are $\Delta^{\mu\nu}_\pm$
given by
(\ref{eq:dteta}) and (\ref{eq:delta}).
Using these laws one can show that
the equation of motion of the original theory transform
to  an identity (Bianchi identity) in the T-dual theory, and vice verse.
From
(\ref{eq:zapr}) and (\ref{eq:zakon}),
using (\ref{eq:xV})  one obtains
\begin{equation}
\beta^\pm_\mu(x)=\mp\big({^\diamond\Pi}_{\pm\mu\nu}(x)-\Pi_{\pm\mu\nu}(x)\big)\partial_\mp x^\nu.
\end{equation}

In section \ref{sec:dualdual}, the T-dual theory was transformed to the original theory.
Comparing the solutions for the auxiliary fields we obtain the following
T-dual coordinate transformation law
\begin{equation}\label{eq:drza}
\kappa\partial_{\pm}y_\mu\cong -2\kappa\,^\dag\Pi_{\mp\mu\nu}(z)\Big(
\partial_{\pm}z^\nu
\pm\kappa\,^\star\beta^{\mp\nu}(z)\Big).
\end{equation}
In the first order this law implies
\begin{equation}
y^{(1)}_\mu\cong
U^{(1)}_\mu(z).
\end{equation}
Substituting
the explicit value of the dual beta function
(\ref{eq:veza}),
we obtain
\begin{equation}\label{eq:zdr}
\partial_{\pm}y_\mu\cong
-2\,^\diamond\Pi_{\mp\mu\nu}(z)
\partial_{\pm}z^\nu,
\end{equation}
with $^\diamond\Pi_{\mp\mu\nu}$ defined in (\ref{eq:dpi}).
Eliminating 
$\partial_{\pm}z^\mu$
from (\ref{eq:drza}) and (\ref{eq:zdr}), using (\ref{eq:xV}) one obtains
\begin{equation}
^\star\beta^{\pm\mu}
\big(V(y)\big)
=\mp\Big({^\diamond\Theta}^{\mu\nu}_\mp
-\Theta^{\mu\nu}_\mp
+2\Delta^{\mu\nu}_\mp\Big)\partial_\pm y_\nu.
\end{equation}

Let us show that
the T-dual coordinate transformation laws (\ref{eq:zapr}) and (\ref{eq:drza}) are inverse to each other.
Multiplying (\ref{eq:drza}) by $^\dag\Theta_{\pm}(V(y))\cong{^\dag\Theta}_{\pm}(z)$
we obtain
\begin{equation}
\kappa\,^\dag\Theta_{\pm}^{\mu\nu}(V(y))
\partial_{\pm}y_\nu\cong
-\partial_{\pm}z^\mu
\mp\kappa\,{^\star\beta}^{\mp\mu}(z).
\end{equation}
Using (\ref{eq:veza}), (\ref{eq:dagteta}) and
 (\ref{eq:dteta}) it becomes
\begin{eqnarray}
\partial_{\pm}z^\mu&\cong&
-\kappa\,^\dag\Theta_{\pm}^{\mu\nu}(V(y))
\partial_{\pm}y_\nu
\mp2\kappa
\,{^\diamond\Theta}^{(1)\mu\nu}_\pm(V(y))\beta^{\mp}_\nu(V(y))
\nonumber\\
&=&
 -\kappa\,{\Theta}^{\mu\nu}_{\pm}\big( V(y)\big)
\Big{[}
\partial_{\pm}y_\nu
\pm2\beta^{\mp}_\nu\big( V(y)\big)
\Big{]}
\nonumber\\
&&+\kappa\Delta^{\mu\nu}_{\pm}(V(y))
\partial_{\pm}y_\nu
\mp\kappa
\,{\Theta}^{\mu\nu}_{1\pm}(V(y))\beta^{\mp}_\nu(V(y)).
\end{eqnarray}
Recalling the definitions (\ref{eq:delta}), (\ref{eq:teta}) and (\ref{eq:betaexp})
the last two terms cancel out and one obtains
\begin{equation}\label{eq:zakondva}
\partial_{\pm}z^\mu\cong
-\kappa\,{\Theta}^{\mu\nu}_{\pm}\big( V(y)\big)
\Big{[}
\partial_{\pm}y_\nu
\pm2\beta^{\mp}_\nu\big( V(y)\big)
\Big{]},
\end{equation}
which is just (\ref{eq:zapr}).
The equivalent conclusion that
(\ref{eq:zakon}) and (\ref{eq:zdr}) are
inverse to each other,
follows from
(\ref{eq:diaminv}).

Let us finally show,
that using the T-dual coordinate transformation laws one can confirm that
the equations of motion and the Bianchi identities of the original and the T-dual theory interchange.
Applying the transformation law (\ref{eq:zdr}) to
the identity
\begin{equation}
\partial_{+}\partial_{-}y_\mu-
\partial_{-}\partial_{+}y_\mu=0,
\end{equation}
we obtain
\begin{equation}\label{eq:jk}
\partial_{+}\partial_{-}z^\mu
+(^\diamond\Pi_{+}-{^\diamond\Pi}_{-})^{-1\mu\nu}
\Big(
\partial_\rho{^\diamond\Pi}_{+\nu\sigma}
-\partial_\sigma{^\diamond\Pi}_{-\nu\rho}
\Big)
\partial_{+}z^\rho
\partial_{-}z^\sigma=0.
\end{equation}
Using the explicit form of the composition ${^\diamond\Pi}_{\pm\mu\nu}$, given by  (\ref{eq:dpi}),
expression (\ref{eq:razlika}) and the value of the Christoffel symbol
 (\ref{eq:chris}), we obtain
\begin{equation}
(^\diamond\Pi_{+}-{^\diamond\Pi}_{-})^{-1\mu\nu}
\Big(
\partial_\rho{^\diamond\Pi}_{+\nu\sigma}
-\partial_\sigma{^\diamond\Pi}_{-\nu\rho}
\Big)=
-B^\mu_{\ \rho\sigma}+\Gamma^\mu_{\rho\sigma}.
\end{equation}
So, (\ref{eq:jk}) is the initial equation of motion (\ref{eq:motion1}).

The equation of motion of the T-dual theory (\ref{eq:tdualaction1}) is
\begin{equation}\label{eq:djkb}
\partial_{+}\Big[\big(
{^\diamond\Theta}_{-}
+\Delta_{-}\big)^{\mu\nu}\big(V(y)\big)
\partial_{-}y_\nu
\Big]
-\partial_{-}\Big[\big(
{^\diamond\Theta}_{+}
+\Delta_{+}
\big)^{\mu\nu}\big(V(y)\big)
\partial_{+}y_\nu
\Big]
=0.
\end{equation}
Using the T-dual coordinate transformation law (\ref{eq:zdr}) (with $z^\mu=x^\mu$) and (\ref{eq:xV}),
we obtain
\begin{equation}\label{eq:djkb}
\partial_{+}\Big[\big(
{^\diamond\Theta}_{-}
+\Delta_{-}\big)^{\mu\nu}\big(x\big)
{^\diamond\Pi}_{+\nu\rho}(x)
\partial_{-}x^\rho
\Big]
-\partial_{-}\Big[\big(
{^\diamond\Theta}_{+}
+\Delta_{+}
\big)^{\mu\nu}\big(x\big)
{^\diamond\Pi}_{-\nu\rho}(x)
\partial_{+}x^\rho
\Big]
=0,
\end{equation}
which with a help of
(\ref{eq:diaminv}) is just the identity
\begin{equation}
\partial_{+}\partial_{-}x^\mu-
\partial_{-}\partial_{+}x^\mu=0.
\end{equation}

\section{Original and dual geometries}\label{sec:uporedjenje}
\cleq

Let us discuss what the geometry of the T-dual  theory looks like and compare it with the geometry of the original theory. To simplify discussion we will put the constant part of the original Kalb-Ramond field, $b_{\mu \nu}$, to zero. In fact it appears in front of a topological term and can contribute only in the quantum theory.

Let us first note the substantial difference  between our T-dual theories and the standard $\sigma$-formulations of string theories. In our approach argument of the background fields is expression $V^\mu$, which is a line integral of
the T-dual coordinate derivatives. For $b_{\mu \nu}=0$ it essentially depends on  a double coordinate ${\tilde{y}}_\mu\equiv \int_{P} d\xi^\alpha \varepsilon^\beta_{\ \alpha}\partial_{\beta}y_\mu$, eq.(\ref{eq:y}), which makes T-dual theory non-geometric. In some particular examples such theories are known as the theories with R-flux. For these theories,
the equations of motion are not necessarily equal to the standardly
derived space-time equations of motion for the background fields depending just on the coordinate $x^\mu$ or $y_\mu$. Although we do not expect that all relations between background fields of the original and T-dual theories in our approach will coincide with those in the literature, we are going to compare them.

Let us first make a simple qualitative analysis. In the approximation of the first order it is easy to see, without calculation, that the T-dual space-time equations of motion are satisfied. In fact
both the dual metric ${}^\star G^{\mu\nu}$ and the dual Kalb-Ramond field ${}^\star B^{\mu \nu}$ are linear in coordinates with infinitesimal coefficients. Consequently,
dual Christoffel  ${}^\star \Gamma_\mu^{\nu \rho}$ and dual field strength  ${}^\star B^{\mu \nu \rho}$ are constant and infinitesimal.
So, both dual space-time equations, for the metric and for the Kalb-Ramond field,  are equal to the second order infinitesimals which are neglected, meaning they are
satisfied.
In our case of the second order approximation, a similar analysis shows that both Riemann tensor and the square of the Kalb-Ramond field are constant and the second order infinitesimals in both initial and T-dual theories.

In this section we are going to discuss the following issues: the geometries and the space-time equations of motion of the initial and T-dual theories as well as the integrability conditions of Ref.
\cite{ALV}.

\subsection{Geometry of the original theory}

For the original theory we take
\begin{eqnarray}
G_{\mu\nu}(x)=g_{\mu\nu}+3h^{2}_{\mu\nu}(x) \, , \qquad  B_{\mu\nu}(x)= h_{\mu\nu}(x) \, ,
\end{eqnarray}
so that  the corresponding Christoffel symbols are linear in coordinate and infinitesimals of the second order while the Kalb-Ramond field strength is a constant  infinitesimal of the first order
\begin{equation}
\Gamma^\mu_{\nu\rho}=-
B^{\mu}_{\ \nu\sigma}h^{\sigma}_{\ \rho}
-B^{\mu}_{\ \rho\sigma}h^{\sigma}_{\ \nu} \, , \qquad H_{\mu \nu \rho} = B_{\mu \nu \rho} \, .
\end{equation}
Therefore, the Riemann tensor is  a constant infinitesimal of the second  order
\begin{equation}
R^\rho{}_{\sigma\mu\nu}= \frac{2}{3}B_{\mu\nu}^{\ \ \,\varepsilon}B_{\varepsilon\ \sigma}^{\ \rho}
+\frac{1}{3}  (B^\rho_{\ \mu\varepsilon}B^\varepsilon_{\ \sigma\nu} - B_{\sigma\mu}^{\ \ \,\varepsilon}B_{\varepsilon\ \nu}^{\ \rho} ),
\end{equation}
which produces the constant second order infinitesimal Ricci tensor
\begin{equation}
R_{\mu\nu}=B_{\mu\rho\sigma}B^{\rho\sigma}_{\ \ \nu}.
\end{equation}
Note that the covariant derivative of the field strength is equal to the ordinary derivative $D_\mu H^\mu{}_{ \rho \sigma} = \partial_\mu H^\mu {}_{\rho \sigma}$.
This is the consequence of the fact that the Christoffel symbols are infinitesimals of the second order and the terms $\Gamma H$ are infinitesimals of the third order which should be neglect in our case. When  $H^\mu{}_{ \rho \sigma}$ is constant one has $D_\mu H^\mu{}_{ \rho \sigma} =0$.

Therefore, the space-time equations of motion can be  written in the form
\begin{eqnarray}
S_{\mu\nu} = 0 \,  , \qquad  D_\mu H^\mu{}_{ \rho \sigma} =0 \, ,
\end{eqnarray}
where for future benefits, following Ref.\cite{ALV}, we introduced the tensors
\begin{equation}
S^\rho{}_{\sigma\mu\nu}  = R^\rho{}_{\sigma\mu\nu} -\frac{2}{3}B_{\mu\nu}^{\ \ \,\varepsilon}B_{\varepsilon\ \sigma}^{\ \rho}
-\frac{1}{3} ( B^\rho_{\ \mu\varepsilon}B^\varepsilon_{\ \sigma\nu} - B_{\sigma\mu}^{\ \ \,\varepsilon}B_{\varepsilon\ \nu}^{\ \rho}    )\, ,
\end{equation}
and
\begin{eqnarray}
S_{\mu\nu} = S^\rho{}_{\mu \rho \nu}  \,  .
\end{eqnarray}
Note that the coefficients in front of the squares of field strength differ from that in Ref.\cite{ALV}, because of a different notation. In both articles they are adjusted in such a way that $S_{\mu\nu} = 0$ is  the space-time equation of motion.

\subsection{Geometry of the T-dual theory}

The background fields of the T-dual theory are
\begin{eqnarray}
{}^\star G^{\mu\nu} = (g^{-1})^{\mu\nu}   \, , \qquad  {}^\star B^{\mu\nu}(V)= - (g^{-1}hg^{-1})^{\mu\nu}(V) \, ,
\end{eqnarray}
where
\begin{equation}
V^\mu=(g^{-1})^{\mu\nu}\tilde{y}_\nu.
\end{equation}
In comparison to the
original theory, the term $h^2$ of the metric tensor is missing and the background fields depend on $V^\mu$ instead of $x^\mu$.

The dual metric is constant and therefore the dual Christoffel symbol, dual Riemann and Ricci tensors are zero.
But, the dual Kalb-Ramond field strength is constant infinitesimal of the first order  ${}^\star H_{\mu\nu\rho} = - B_{\mu\nu\rho}$.
As we explained in the beginning of this section the T-dual fields do not satisfy the standard space-time equations of motion because the space is nongeometric and  the background fields depend on the dual coordinate $\tilde{y}$.
The T-dual space-time equations of motion are 
\begin{eqnarray}
{}^\star R^{\mu\nu} = 0 \,  , \qquad  {}^\star D^\mu {}^\star H_\mu{}^{ \rho \sigma} =0 \, ,
\end{eqnarray}
where again dual covariant derivative is equal to the ordinary derivative. It is interesting to note that although the initial theory is curved, the corresponding T-dual is flat. It seems that at least in the second order the T-duality acts as a parallelizable transformation. This assumption should be checked
in the higher orders of approximation.

\subsection{Relation with Ref.\cite{ALV}}

Although, as we explained, we should not expect for our background to satisfy the pseudoduality conditions of Ref.\cite{ALV}, we are going to discuss  the relation with this article. Let us first note that $\beta^\pm_\mu$ functions,  introduced in the generalized T-dualization procedure, 
originate from the fact
that T-dual background fields do not depend on the coordinate $y_\mu$ but on its dual $V^\mu=(g^{-1})^{\mu\nu}\tilde{y}_\nu$. So, in order to compare our relations with the conditions derived in the literature we will omit the term $\beta^-_\mu \Theta^{\mu\nu}_- \beta^+_\nu$. Then, the T-dual metric tensor acquires  the quadratic term and the Kalb-Ramond field is unchanged 
\begin{eqnarray}\label{eq:gbbz}
{}^\star G^{\mu\nu} = (g^{-1})^{\mu\nu}  + (h^2)^{\mu\nu}(V)  \, , \qquad  {}^\star B^{\mu\nu}(V)= - (g^{-1}hg^{-1})^{\mu\nu}(V) \,.
\end{eqnarray}

Secondly, one can note that the pseudoduality relation in Ref.\cite{ALV}, which is the starting point in that paper corresponds to
the relation (\ref{eq:zdr}) in this paper.
Taking $b_{\mu \nu}=0$ the relation (\ref{eq:zdr}) reduces to
\begin{equation}\label{eq:zdrb}
\partial_{\pm}y_\mu\cong \mp (g \mp 3h + 6 h^2)_{\mu\nu}(x)\partial_{\pm} x^\nu  \equiv \pm T_\pm \partial_{\pm} x^\nu   \, ,
\end{equation}
where $T_\pm$ is the notation from the  article \cite{ALV} where only the case $T_+ = T_-$ was treated.
Obviously, the T-dual coordinate transformation laws differ at least by
the term $3 h$.

The Christoffel symbol and the Kalb-Ramond field strength for the background fields (\ref{eq:gbbz}) are
\begin{equation}
^\star\Gamma_\mu^{\nu\rho}=-\frac{1}{3}\Big(
B_{\mu\ \sigma}^{\ \nu}h^{\sigma\rho}
+B_{\mu\ \sigma}^{\ \rho}h^{\sigma\nu}
\Big),\quad {^\star H}^{\mu\nu\rho}= -B^{\mu\nu\rho}.
\end{equation}
The Riemann tensor is
\begin{equation}
^\star R_\rho^{\mu\sigma\nu}=
-\frac{2}{9}B_{\rho\ \varepsilon}^{\ \mu}B^{\varepsilon\nu\sigma}
-\frac{1}{9}B_{\rho\ \varepsilon}^{\ \nu}B^{\varepsilon\mu\sigma}
+\frac{1}{9}B_{\rho\ \varepsilon}^{\ \sigma}B^{\varepsilon\mu\nu},
\end{equation}
and the Ricci tensor equals
\begin{equation}
^\star R^{\mu\nu}=\frac{1}{3}B^{\mu\varepsilon\rho}B_{\varepsilon\rho}^{\ \ \nu}.
\end{equation}

Note that the Christoffel symbols,  Riemann and Ricci tensors are one third of the corresponding variables of the original theory.
The same as 
in the original theory, the Christoffel symbols are infinitesimals of the second order and the covariant derivative of the field strength is equal to the ordinary derivative ${}^\star D^\mu {}^\star H^{\mu \rho \sigma} = \partial^\mu {}^\star H^{\mu \rho \sigma}$. Because the tensor ${}^\star H^{\mu\rho \sigma}$ is constant the right hand side is zero. 

The dual space-time equations of motion can be written as
\begin{eqnarray}
{}^\star S^{\mu\nu} = 0 \,  , \qquad {}^\star  D^\mu {}^\star H_\mu{}^{ \rho \sigma} =0 \, ,
\end{eqnarray} 
where we define the dual  tensors ${}^\star S_\rho{}^{\sigma\mu\nu}$ with an additional coefficient $\frac{1}{3}$ in the last three terms in comparison to the tensor $ S^\rho{}_{\sigma\mu\nu}$ 
\begin{equation}
{}^\star S_\rho{}^{\sigma\mu\nu}  = {}^\star R_\rho{}^{\sigma\mu\nu} - \frac{2}{9} B^{\mu\nu}_{\ \ \,\varepsilon}B^{\varepsilon\ \sigma}_{\ \rho}
-\frac{1}{9}  (B_\rho^{\ \mu\varepsilon}B_\varepsilon^{\ \sigma\nu} - B^{\sigma\mu}_{\ \ \,\varepsilon} B^{\varepsilon\ \nu}_{\ \rho} ) \, ,
\end{equation}
and as usual 
\begin{eqnarray}
{}^\star S^{\mu\nu} = {}^\star  S_\rho{}^{\mu \rho \nu}  \,  .
\end{eqnarray}

The pseudoduality conditions of the Ref. \cite{ALV}, in our notation read as follows 
\begin{eqnarray}
{}^\star S^\rho{}_{\sigma\mu\nu}   =  -  S^\rho{}_{\sigma\mu\nu} \,  , \qquad {}^\star  D_\mu {}^\star H^\mu{}_{ \rho \sigma} = - D_\mu  H^\mu{}_{ \rho \sigma} \, .
\end{eqnarray}
Note that because both equations are infinitesimal we can raise and lower indices with the constant part of the metric. Both  pseudoduality conditions are fulfilled because all terms are separately equal to zero.  
The second relation is valid without derivatives as well ${}^\star H^\mu{}_{ \rho \sigma} = -  H^\mu{}_{ \rho \sigma}$.

\section{Conclusion}
\cleq
In this paper,
we presented the T-dualization procedure applicable to string backgrounds
with nontrivial Ricci tensor and
 without isometries.
The procedure is the generalization of the one given in paper \cite{DS},
for a weakly curved background.
It was applied to a string moving in the weakly curved background of the second order,
composed of a linearly coordinate dependent  Kalb-Ramond field with an infinitesimal strength
and a metric with an infinitesimal of the second order quadratic in coordinate term.

The generalized Buscher procedure was not applicable
to the second order weakly curved background, because the action does not possess a global symmetry.
If there is no global symmetry,
 there is no corresponding gauge symmetry, which is the crucial ingredient of the T-dualization procedure.
However,
it is possible to construct an
 auxiliary action, which plays the role of the gauge fixed action.
The auxiliary action is constructed from the initial action,
substituting the derivatives of the coordinates by
some auxiliary fields, and the background fields argument by
a line integral of these auxiliary fields.
This action
reduces to the initial action and to the T-daul action on its equations of motion.
So, there is a full analogy between the T-dualization procedures for backgrounds
with and without a global symmetry.
The only difference is in fact the interpretation of the auxiliary fields which are understood
as  the gauge fields in the case of a background with symmetry.

The realization of the new generalized Buscher procedure is more complicated.
The main problem is to solve the equations of motion, obtained varying the auxiliary action
with respect to the
auxiliary fields, in terms of the Lagrange multipliers.
To solve them, one should iteratively calculate the argument of the background fields and  the beta functions defined in Apps. \ref{sec:betaj} and \ref{sec:betad}.
It turns out that
the argument of the dual background fields has a more complicated form then in the
weakly curved background of the first order.
This argument represents
the complicated structure of the dual geometry.
In the first order the argument is given in terms of the dual coordinate $y_\mu$ and its double $\tilde{y}_\mu$.
In the second order the argument is given in terms of
the dual coordinate, its double and an additional form
$M_{\pm\mu\nu}(y)\equiv
\frac{1}{2}
\int d\xi^\pm\,\Big( y^{(0)}_{\pm\mu}\partial_\pm y^{(0)}_{\pm\nu}
-y^{(0)}_{\pm\nu}\partial_\pm y^{(0)}_{\pm\mu}
\Big)$,
which can be interpreted as the left and the right "angular momentum".

 Applying the T-dualization procedure to all the coordinates of the initial theory, the theory transforms to a T-dual theory.
The initial background which is curved and geometric transforms to a non geometric curved background
$$
\Pi_{+\mu\nu}(x)\rightarrow
\,^\star\Pi_{+}^{\mu\nu}(y)=
\frac{\kappa}{2}\,^\dag\Theta_{-}^{\mu\nu}\big(V(y)\big),
$$
where  $\Theta_{-}^{\mu\nu}$ is defined in (\ref{eq:tetadualna}).
Consequently the T-dual theory can not be directly compared with the standard  theories,
 where the background fields depend on the ordinary coordinates.
The T-dual Riemann tensor is zero, which means that T-dual background is flat.
It would be interesting to check whether T-duality 
in the higher orders  can make the target  space parallelizable. 
Although the T-dual background is flat,
 certain adjustments can be made in order for our approximation
to satisfy the relations analogue to 
the general relations of Ref. \cite{ALV}.

Applying the procedure to all the dual coordinates, T-dual theory transforms to
the initial theory.
Comparing the solutions for the auxiliary fields we obtain
the T-dual coordinate transformation laws,
connecting initial and dual coordinates,
which are inverse to each other.
Using these laws one confirms that the equations of motion and the Bianchi identity of one theory transform to the Bianchi identity and the equations of motion of the other theory.
Because these laws have obtained the second order correction,
they will enable further investigation of the non-commutativity properties
of the spaces connected by T-duality.
Furthermore,
the laws are the basis for a double formulation \cite{D,S1,S2,NS}
where T-duality is
interpreted as an exchange of the initial and dual coordinates.

So, we showed that the T-dualization of the theory with a non-trivial Ricci tensor
and without global symmetry is possible,
and that  it does not break the standard features of T-duality.
The non geometric structure  of T-dual theory is much richer than in the cases previously analyzed
 and may be a subject of further investigations.

\appendix
\section{The expansion of the background fields}
\cleq

All the expressions will be divided into its zeroth, first and second order values,
for example $G_{\mu\nu}=G_{0\mu\nu}+G_{1\mu\nu}+G_{2\mu\nu}$.
By
\begin{equation}
G^{(1)}_{\mu\nu}=G_{0\mu\nu}+G_{1\mu\nu},
\end{equation}
we mark the value up to the first order.
The inverse of $G_{\mu\nu}$ is given by
\begin{equation}
(G^{-1})^{\mu\nu}=
(G^{-1})_{0}^{\mu\nu}-
\Big{[}
(G^{-1})_{0}\Big{(}
G_{1}+G_{2}-G_{1}(G^{-1})_{0}G_{1}
\Big{)}(G^{-1})_{0}
\Big{]}^{\mu\nu}
\end{equation}

\begin{itemize}
\item Original background fields
\begin{eqnarray}\label{eq:gorg}
&\quad G_{0\mu\nu}=g_{\mu\nu},&\quad B_{0\mu\nu}=b_{\mu\nu},\quad  \Pi_{0\pm\mu\nu}=b_{\mu\nu}\pm\frac{1}{2}g_{\mu\nu},
\nonumber\\
&G_{1\mu\nu}=0,&\quad B_{1\mu\nu}=h_{\mu\nu},\quad
\Pi_{1\pm\mu\nu}=h_{\mu\nu},
\nonumber\\
&\quad G_{2\mu\nu}=3h^{2}_{\mu\nu},&\quad B_{2\mu\nu}=0,\qquad
\Pi_{2\pm\mu\nu}=\pm\frac{3}{2}h^{2}_{\mu\nu}.
\end{eqnarray}

\item Inverse of a metric $(G^{-1})^{\mu\nu}$
\begin{eqnarray}
(G^{-1})_{0}^{\mu\nu}&=&({g}^{-1})^{\mu\nu},
\nonumber\\
(G^{-1})_{1}^{\mu\nu}&=&0,
\nonumber\\
(G^{-1})_{2}^{\mu\nu}&=&-3
(g^{-1}h^{2}g^{-1})^{\mu\nu}.
\end{eqnarray}

\item Effective metric $(G_{E})_{\mu\nu}=G_{\mu\nu}-4(BG^{-1}B)_{\mu\nu}$
\begin{eqnarray}
(G_{E0})_{\mu\nu}&=&g_{\mu\nu}-4b^{2}_{\mu\nu}=(g_{E})_{\mu\nu},
\nonumber\\
(G_{E1})_{\mu\nu}&=&-4(bh+hb)_{\mu\nu},
\nonumber\\
(G_{E2})_{\mu\nu}&=&-h^{2}_{\mu\nu}
+12(bh^{2}b)_{\mu\nu}.
\end{eqnarray}

\item Effective metric inverse
 \begin{eqnarray}\label{eq:invmet}
(G_{E}^{-1})_{0}^{\mu\nu}&=&(g_{E}^{-1})^{\mu\nu},
\nonumber\\
(G_{E}^{-1})_{1}^{\mu\nu}&=&4\Big{[}g_{E}^{-1}\Big{(}
bh+hb\Big{)}g_{E}^{-1}\Big{]}^{\mu\nu}
=-2\kappa(\theta_{0}hg_{E}^{-1}+g_{E}^{-1}h\theta_{0})^{\mu\nu},
\nonumber\\
(G_{E}^{-1})_{2}^{\mu\nu}&=&
-\Big{[}g_{E}^{-1}\Big{(}
-h^{2}
+12bh^{2}b
-16(bh+hb)g_{E}^{-1}(bh+hb)
\Big{)}
g_{E}^{-1}\Big{]}^{\mu\nu}
\nonumber\\
&=&-3(g_{E}^{-1}h^{2}g_{E}^{-1})^{\mu\nu}
+6\kappa^{2}(\theta_{0}h^{2}\theta_{0})^{\mu\nu}
+4(g_{E}^{-1}hg_{E}^{-1}hg_{E}^{-1})^{\mu\nu}
\nonumber\\
&+&4\kappa^{2}(\theta_{0}hg_{E}^{-1}h\theta_{0}
+\theta_{0}hg_{E}^{-1}h\theta_{0}
+g_{E}^{-1}h\theta_{0}h\theta_{0})^{\mu\nu}.
\end{eqnarray}

\item Parameter of noncommutativity $\theta^{\mu\nu}=-\frac{2}{\kappa}(G^{-1}_{E}BG^{-1})^{\mu\nu}$
\begin{eqnarray}\label{eq:tetaraz}
\theta^{\mu\nu}_{0}&=&-\frac{2}{\kappa}(g^{-1}_{E}bg^{-1})^{\mu\nu},
\nonumber\\
\theta^{\mu\nu}_{1}&=&-\frac{2}{\kappa}(g^{-1}_{E}(h+4bhb)g^{-1}_{E})^{\mu\nu}
=-\frac{2}{\kappa}(g^{-1}_{E}hg^{-1}_{E}+\kappa^{2}\theta_{0}h\theta_{0})^{\mu\nu},
\nonumber\\
\theta^{\mu\nu}_{2}&=&
-3(\theta_{0}h^{2}g^{-1}_{E}+g^{-1}_{E}h^{2}\theta_{0})^{\mu\nu}
+4\kappa^{2}(\theta_{0}h\theta_{0}h\theta_{0})^{\mu\nu}
\nonumber\\
&&
+4(\theta_{0}hg^{-1}_{E}hg^{-1}_{E}
+g^{-1}_{E}h\theta_{0}hg^{-1}_{E}
+g^{-1}_{E}hg^{-1}_{E}h\theta_{0}
)^{\mu\nu}.
\end{eqnarray}

\item Theta function $\Theta^{\mu\nu}_{\pm}=-\frac{2}{\kappa}(G^{-1}_{E}\Pi_{\pm}G^{-1})^{\mu\nu}
=\theta^{\mu\nu}\mp\frac{1}{\kappa}(G^{-1}_{E})^{\mu\nu}$
\begin{eqnarray}\label{eq:teta}
\Theta^{\mu\nu}_{0\pm}&=&
\theta^{\mu\nu}_{0}\mp\frac{1}{\kappa}(g^{-1}_{E})^{\mu\nu},
\nonumber\\
\Theta^{\mu\nu}_{1\pm}&=&
-2\kappa\Big{[}
\Theta_{0\pm}h\Theta_{0\pm}
\Big{]}^{\mu\nu},
\nonumber\\
\Theta^{\mu\nu}_{2\pm}&=&
\Theta_{0\pm}^{\mu\rho}\Big{[}
\pm3\kappa h^{2}
+4\kappa^{2}h\Theta_{0\pm}h
\Big{]}_{\rho\sigma}\Theta^{\sigma\nu}_{0\pm}
\nonumber\\
&=&\pm3\kappa
\Theta_{0\pm}^{\mu\rho}h^{2}_{\rho\sigma}\Theta^{\sigma\nu}_{0\pm}
+4\Delta^{\mu\nu}_\pm.
\end{eqnarray}

\item The second order
contribution
$\partial_{+}y_\mu\,\Delta^{\mu\nu}_{-}\big(V_{0}(y)\big)
\partial_{-}y_\nu
\equiv-
4\beta^{-}_{1\mu}\big(V_{0}(y)\big)
\Theta_{0-}^{\mu\nu}\beta^{+}_{1\nu}\big(V_{0}(y)\big)
$

\begin{equation}\label{eq:delta}
\Delta^{\mu\nu}_\pm=\frac{\kappa}{2}
\Theta_{1\pm}^{\mu\rho}\Pi_{0\mp\rho\sigma}\Theta_{1\pm}^{\sigma\nu}
=
\kappa^{2}
\Big(
\Theta_{0\pm}h\Theta_{0\pm}h\Theta_{0\pm}
\Big)^{\mu\nu}.
\end{equation}

\item Dual background fields composition $^\dag\Theta_\pm^{\mu\nu}=\Theta_\pm^{\mu\nu}
-\Delta^{\mu\nu}_\pm
$
\begin{eqnarray}\label{eq:dagteta}
^\dag\Theta_{0\pm}^{\mu\nu}&=&\Theta_{0\pm}^{\mu\nu}\,,
\nonumber\\
^\dag\Theta_{1\pm}^{\mu\nu}&=&\Theta_{1\pm}^{\mu\nu}\,,
\nonumber\\
^\dag\Theta_{2\pm}^{\mu\nu}&=&
3\kappa^{2}\Big[\Theta_{0\pm}h
\Big(
\theta_{0}
\pm\frac{1}{\kappa}(g^{-1}-g^{-1}_{E})
\Big)
h\Theta_{0\pm}\Big]^{\mu\nu}.
\end{eqnarray}

\item Dual background fields composition $^\dag\Pi_{\pm\mu\nu}=\Pi_{\pm\mu\nu}+2\kappa
\Pi_{\pm\mu\rho}\Delta^{\rho\sigma}_\mp\Pi_{\pm\sigma\nu}$
\begin{eqnarray}\label{eq:pik}
^\dag\Pi_{0\pm\mu\nu}&=&\Pi_{0\pm\mu\nu},
\nonumber\\
^\dag\Pi_{1\pm\mu\nu}&=&\Pi_{1\pm\mu\nu},
\nonumber\\
^\dag\Pi_{2\pm\mu\nu}&=&
\Pi_{2\pm\mu\nu}
+\frac{\kappa}{2}h_{\mu\rho}\Theta^{\rho\sigma}_{0\mp}h_{\sigma\nu}.
\end{eqnarray}

The last two compositions are inverse to each other
\begin{equation}\label{eq:dagpt}
^\dag\Pi_{\pm\mu\nu}{^\dag\Theta_\mp^{\nu\rho}}=\frac{1}{2\kappa}
\delta_\mu^\rho.
\end{equation}

\item Functions $^\diamond F_\pm$
\begin{eqnarray}
^\diamond F_\pm&\equiv&
\sum_{n=0}^{2}(1+\frac{n}{2})F_{n\pm},
\\
\label{eq:dteta}
^\diamond\Theta^{\mu\nu}_\pm&\equiv&
\Theta_{0\pm}^{\mu\nu}
+\frac{3}{2}\Theta_{1\pm}^{\mu\nu}
+2\Theta_{2\pm}^{\mu\nu},
\\\label{eq:dpi}
^\diamond\Pi_{\pm\mu\nu}&\equiv&
\Pi_{0\pm\mu\nu}+
\frac{3}{2}\Pi_{1\pm\mu\nu}+
2\Pi_{2\pm\mu\nu}.
\end{eqnarray}

\item Inverses of functions $^\diamond\Theta^{\mu\nu}_\pm$ and $^\diamond\Pi_{\pm\mu\nu}$
\begin{eqnarray}\label{eq:diaminv}
(^\diamond\Pi_{\pm\mu\nu}
+2\kappa\Pi_{0\pm\mu\rho}\Delta^{\rho\sigma}_\mp\Pi_{0\pm\sigma\nu})
\,^\diamond\Theta^{\nu\rho}_\mp=\frac{1}{2\kappa}\delta_\mu^\rho,
\nonumber\\
^\diamond\Pi_{\pm\mu\nu}(\,
^\diamond\Theta^{\nu\rho}_\mp+\Delta_\mp^{\nu\rho})=
\frac{1}{2\kappa}\delta_\mu^\rho.
\end{eqnarray}
In the first order one has
\begin{equation}\label{eq:prr}
^\diamond\Pi_{\pm\mu\nu}^{(1)}{
^\diamond\Theta^{(1)\nu\rho}_\mp}=
\frac{1}{2\kappa}\delta_\mu^\rho.
\end{equation}
\item Difference
\begin{equation}^\diamond\Pi_{+\mu\nu}-{^\diamond\Pi}_{-\mu\nu}=
g_{\mu\nu}+6h^{2}_{\mu\nu},
\end{equation}
and its inverse
\begin{equation}\label{eq:razlika}
(^\diamond\Pi_{+\mu\nu}-{^\diamond\Pi}_{-\mu\nu})^{-1}=
(g^{-1})^{\mu\nu}-6(h^{2})^{\mu\nu}.
\end{equation}
\end{itemize}

\section{Beta function $\beta^\pm_\mu$}\label{sec:betaj}
\cleq

Let us calculate the beta functions defined in (\ref{eq:betadef}), for the action (\ref{eq:spom}).
The variation of the action over the background fields argument is
\begin{equation}
\delta_{V}S_{aux}=\kappa\int d^{2}\xi
\Big[
\varepsilon^{\alpha\beta}\partial_\rho B_{\mu\nu}
+\frac{1}{2}\eta^{\alpha\beta}\partial_\rho G_{\mu\nu}
\Big]\partial_\alpha V^\mu\partial_\beta V^\nu\delta V^\rho.
\end{equation}
Partially integrating, using the zeroth order equation of motion $\eta^{\alpha\beta}\partial_\alpha\partial_\beta V^\mu=0$ in a quadratic in $B_{\mu\nu\rho}$ terms, we obtain
\begin{eqnarray}
\delta_{V}S_{aux}&=&-\kappa\int d^{2}\xi\Big[
\Big(
\varepsilon^{\alpha\beta}\partial_\rho B_{\mu\nu}
+\frac{1}{2}\eta^{\alpha\beta}\partial_\rho G_{\mu\nu}
\Big)V^\mu\partial_\beta V^\nu\delta v^\rho_\alpha
\nonumber\\
&+&\frac{1}{2}\eta^{\alpha\beta}\partial_\alpha\partial_\rho G_{\mu\nu}V^\mu\partial_\beta V^\nu\delta V^\rho
\Big].
\end{eqnarray}
Using the explicit value of the initial metric
the second term can be rewritten as
\begin{equation}\label{eq:pomoc}
\frac{1}{2}\eta^{\alpha\beta}\partial_\alpha\partial_\rho G_{\mu\nu}V^\mu\partial_\beta V^\nu\delta V^\rho
=\frac{1}{2}\eta^{\alpha\beta}\partial_\alpha G_{2\mu\rho}(V)\partial_\beta V^\mu
\delta V^\rho
=\frac{1}{2}\eta^{\alpha\beta}\partial_\alpha
\Big(
 G_{2\mu\rho}(V)\partial_\beta V^\mu\Big)
\delta V^\rho,
\end{equation}
and therefore
\begin{equation}
\delta_{V}S_{aux}=-\kappa\int d^{2}\xi\Big[
\Big(
\varepsilon^{\alpha\beta}\partial_\rho B_{\mu\nu}
+\frac{1}{2}\eta^{\alpha\beta}\partial_\rho G_{\mu\nu}
\Big)V^\mu\partial_\beta V^\nu\delta v^\rho_\alpha
-\frac{1}{2}\eta^{\alpha\beta}
 G_{2\mu\rho}(V)\partial_\beta V^\mu
\delta v_\alpha^\rho\Big].
\end{equation}
So, the beta functions, defined by (\ref{eq:betadef}), are
\begin{eqnarray}
\beta^\alpha_\rho&=&
\varepsilon^{\alpha\beta}\partial_\rho B_{\mu\nu}
V^\mu\partial_\beta V^\nu
+\frac{1}{2}\eta^{\alpha\beta}
\Big(
\partial_\rho G_{\mu\nu}(V)V^\mu
-G_{2\nu\rho}(V)
\Big)\partial_\beta V^\nu
\nonumber\\
&=&
-\Big{(}
\epsilon^{\alpha\beta}h_{\mu\nu}(V)
+3\eta^{\alpha\beta}h^{2}_{\mu\nu}(V)
\Big{)}\partial_\beta V^\nu,
\end{eqnarray}
and in the light-cone coordinates they become
\begin{equation}\label{eq:betaexp}
\beta^\pm_\rho(V)=\frac{1}{2}\Big(
\beta^{0}_\rho\pm\beta^{1}_\rho
\Big)=
\frac{1}{2}\Big{(}
\mp h_{\rho\nu}(V)-3h^{2}_{\rho\nu}(V)
\Big{)}\partial_{\mp} V^\nu.
\end{equation}

\section{Dual beta function $^\star\beta^{\pm\mu}$}\label{sec:betad}
\cleq

In this section we will find the beta functions for the dual theory auxiliary action (\ref{eq:sdualaux})
\begin{equation}
^\star S_{aux}=\frac{\kappa^{2}}{2}\int d^{2}\xi\Big{[}
u_{+\mu}\,^\dag\Theta_{-}^{\mu\nu}\big(V(U)\big)\,u_{-\nu}
+\frac{1}{\kappa}(u_{+\mu}\partial_{-}z^\mu
-u_{-\mu}\partial_{+}z^\mu)
\Big{]}.
\end{equation}
We define them as usual by
\begin{equation}\label{eq:betadualdef}
\delta_{V}{^\star S}_{aux}=\frac{\kappa^{2}}{2}\int d^{2}\xi\,
u_{+\mu}\,\partial_{\rho}{}^\dag\Theta_{-}^{\mu\nu}\big(V(U)\big)\,u_{-\nu}
\,\delta V^\rho(U)
=-\frac{\kappa^{2}}{2}\int d^{2}\xi\Big{[}
{^\star\beta}^{+\mu}\delta u_{+\mu}+
^\star\beta^{-\mu}\delta u_{-\mu}\Big{]}.
\end{equation}
Multiplying the equation (\ref{eq:vdt}) by
${^\diamond\Pi}^{(1)}_{\mp\mu\nu}$, defined by (\ref{eq:dpi})
(the inverse of $^\diamond\Theta^{(1)\mu\nu}_\pm$,
see (\ref{eq:prr})), one obtains the
auxiliary fields $u_{\pm\mu}$
in terms of the auxiliary fields $v^\mu_\pm$
\begin{equation}
u_{\pm\mu}^{(1)}=-2{^\diamond\Pi}^{(1)}_{\mp\mu\nu}(V(U))
v_{\pm}^{(1)\mu}(U).
\end{equation}
Substituting these expressions to the first expression in (\ref{eq:betadualdef}) we obtain
\begin{equation}\label{eq:var}
\delta_{V}{^\star S}_{aux}=-2\kappa^{2}\int d^{2}\xi\,
\partial_{+}V^\mu
F_{\mu\nu,\rho}
(V(U))
\partial_{-}V^\nu\delta V^\rho(U),
\end{equation}
with the background field composition
$F_{\mu\nu,\rho}$ defined by
\begin{eqnarray}\label{eq:f}
F_{\mu\nu,\rho}\equiv\Big(
{^\diamond\Pi}_{+}\,
\partial_\rho{^\dag\Theta}_{-}
{^\diamond\Pi}_{+}
\Big)_{\mu\nu}
=-\frac{1}{2\kappa}\partial_{\rho}
\Big{[}
h
+\frac{3}{2}
h\Big(
\kappa\Theta_{0-}+g^{-1}
\Big)h
\Big{]}_{\mu\nu},
\end{eqnarray}
where the second expression is obtained using
(\ref{eq:dpi}) and (\ref{eq:dagteta}).
Partially integrating in (\ref{eq:var}) we obtain
\begin{eqnarray}\label{eq:vari}
\delta_{V}{^\star S}_{aux}&=&\kappa^{2}\int d^{2}\xi\,
\Big{\{}
\Big{[}
V^\mu\partial_{+}
F_{\mu\nu,\rho}(V)
\partial_{-}V^\nu
+
\partial_{+}V^\mu
\partial_{-}
F_{\mu\nu,\rho}(V)
 V^\nu
\Big{]}
\delta V^\rho(U)
\nonumber\\
&+&
V^\mu
F_{\mu\nu,\rho}(V)
\partial_{-}V^\nu\delta v^\rho_{+}(U)
+
\partial_{+}V^\mu
F_{\mu\nu,\rho}(V)
 V^\nu
\delta v^\rho_{-}(U)\Big{\}}.
\end{eqnarray}
The term $
F_{\mu\nu,\rho}(V)
\Big(V^\mu\partial_{+}\partial_{-}V^\nu
+\partial_{+}\partial_{-}V^\mu V^\nu\Big)\delta V^\rho
$
is absent, because, the antisymmetric in first two indices part of $F_{\mu\nu,\rho}$
gives zero, while its symmetric part is of the second order and therefore the whole expression vanishes on the zeroth order equation of motion $\partial_{+}\partial_{-}V^\mu=0$.

Taking  the variation of (\ref{eq:vdt}), using (\ref{eq:dteta}), one obtains
\begin{eqnarray}\label{eq:vprekou}
\delta v^\mu_\pm(U)&=&
-\kappa
^\diamond\Theta^{(1)\mu\nu}_\pm(V(U))\delta u_{\pm\nu}
+\delta v^{\star\mu}_\pm(U),
\\
\delta v^{\star\mu}_\pm(U)&=&
-\frac{3\kappa}{2}\partial_\rho\Theta_{1\pm}^{\mu\nu}u_{0\pm\nu}
\delta V_{0}^\rho(U_{0}).
\end{eqnarray}
Using (\ref{eq:teta}) and (\ref{eq:vo}) one observes that
\begin{equation}
\delta v^{\star\mu}_\pm(U)=
-3\kappa\Theta_{0\pm}^{\mu\nu}\partial_\rho h_{\nu\sigma}\partial_\pm V^\sigma\delta V^\rho.
\end{equation}

Let us calculate the contribution from  $\delta v^{\star\mu}_\pm(U)$ terms.
Because these terms are of the first order it is enough to use the first order value of $F_{\mu\nu,\rho}$, defined by (\ref{eq:f}).
One obtains
\begin{eqnarray}
\delta_{V}{^\star S}_{aux}&=&\kappa^{2}\int d^{2}\xi\,
\Big(
V^\mu
F_{\mu\nu,\rho}(V)
\partial_{-}V^\nu\delta v^{\star\rho}_{+}(U)
+
\partial_{+}V^\mu
F_{\mu\nu,\rho}(V)
 V^\nu
\delta v^{\star\rho}_{-}(U)
\Big)
\nonumber\\
&=&-\frac{3\kappa^{2}}{2}
\int d^{2}\xi\,
\partial_{+}V^\mu\partial_{\rho}
(h\Theta_{0-}h)_{\mu\nu}(V)\partial_{-}V^\nu\delta V^\rho.
\end{eqnarray}
Partially integrating we obtain
\begin{eqnarray}\label{eq:zvezda}
\delta_{V}{^\star S}_{aux}&=&
\frac{3\kappa^{2}}{4}
\int d^{2}\xi\,
\Big[\Big(
V^\mu\partial_{+}\partial_{\rho}(h\Theta_{0-}h)_{\mu\nu}(V)\partial_{-}V^\nu
\nonumber\\
&+&\partial_{+}V^\mu\partial_{-}\partial_{\rho}(h\Theta_{0-}h)_{\mu\nu}(V)V^\nu
\Big)
\delta V^\rho
\nonumber\\
&+&V^\mu\partial_{\rho}(h\Theta_{0-}h)_{\mu\nu}(V)\partial_{-}V^\nu\delta v_{+}^\rho
+\partial_{+}V^\mu\partial_{\rho}(h\Theta_{0-}h)_{\mu\nu}(V)V^\nu\delta v_{-}^\rho
\Big],
\end{eqnarray}
where we again used the zeroth order equation of motion $\partial_{+}\partial_{-}V^\mu=0$.
Finally,
substituting (\ref{eq:vprekou}) and (\ref{eq:f}) into (\ref{eq:vari}),
using (\ref{eq:zvezda}),
and noting that $\partial_\pm\partial_\rho h_{\mu\nu}=0$,
 the variation of the auxiliary action becomes
\begin{eqnarray}
\delta_{V}{^\star S}_{aux}&=&\kappa^{2}\int d^{2}\xi\,
\Big{\{}
\Big{[}
-\frac{3}{4\kappa}V^\mu\partial_{+}\partial_\rho h^{2}_{\mu\nu}(V)
\partial_{-}V^\nu
-\frac{3}{4\kappa}
\partial_{+}V^\mu
\partial_{-}\partial_\rho h^{2}_{\mu\nu}(V)
 V^\nu
\Big{]}
\delta V^\rho(U)
\nonumber\\
&-&
\frac{1}{2\kappa}V^\mu\partial_\rho(
h+\frac{3}{2}h^{2}
)_{\mu\nu}(V)\partial_{-}V^\nu
(-\kappa)
^\diamond\Theta^{(1)\rho\sigma}_{+}(V)\delta u_{+\sigma}
\nonumber\\
&-&
\frac{1}{2\kappa}\partial_{+}V^\mu\partial_\rho(
h+\frac{3}{2}h^{2}
)_{\mu\nu}(V)V^\nu(-\kappa)
^\diamond\Theta^{(1)\rho\sigma}_{-}(V)\delta u_{-\sigma}
\Big{\}}.
\end{eqnarray}

The first two terms can be rewritten as
\begin{eqnarray}
V^\mu\partial_{+}\partial_\rho h^{2}_{\mu\nu}
\partial_{-}V^\nu
+
\partial_{+}V^\mu
\partial_{-}\partial_\rho h^{2}_{\mu\nu}
 V^\nu
&=&
\partial_{+}V^\mu \partial_{-}h^{2}_{\mu\rho}
+\partial_{-}V^\mu \partial_{+}h^{2}_{\mu\rho}
\nonumber\\
&=&\partial_{-}\Big(
\partial_{+}V^\mu h^{2}_{\mu\rho}\Big)
+\partial_{+}\Big(
\partial_{-}V^\mu h^{2}_{\mu\rho}
\Big).
\end{eqnarray}
Partially integrating, using (\ref{eq:vprekou}), we obtain
\begin{eqnarray}
\delta_{V}{^\star S}_{aux}&=&-\frac{\kappa^{2}}{2}\int d^{2}\xi\,
\Big{\{}
\delta u_{-\mu}\,
^\diamond\Theta^{(1)\mu\nu}_{+}(V)
(h-3h^{2})_{\nu\rho}(V)
\partial_{+}V^\rho
\nonumber\\
&+&\delta u_{+\mu}\,
^\diamond\Theta^{(1)\mu\nu}_{-}(V)
(-h-3h^{2})_{\nu\rho}(V)
\partial_{-}V^\rho
\Big{\}}.
\end{eqnarray}

Finally,
recalling (\ref{eq:betaexp})
we obtain the dual beta functions
\begin{equation}\label{eq:veza}
{^\star\beta}^{\pm\mu}(V(U))
=\,
2\,{^\diamond\Theta}^{(1)\mu\nu}_\mp(V(U))\beta^{\pm}_\nu(V(U)).
\end{equation}


\end{document}